# Anomalous Returns in a Neural Network Equity-Ranking Predictor

J.B. Satinover[1] and D. Sornette[2]

[1]Laboratoire de Physique de la Matière Condensée, CNRS UMR6622 and Université des Sciences, Parc Valrose, 06108 Nice Cedex 2, France
[2]Department of Management, Technology and Economics, ETH Zurich, CH-8032, Zurich, Switzerland and Swiss Finance Institute, c/o University of Geneva, 40 blvd. Du Pont d'Arve, CH 1211 Geneva 4, Switzerland

**Abstract**—Using an artificial neural network (ANN), a fixed universe of ~1500 equities from the Value Line index are rank-ordered by their predicted price changes over the next quarter. Inputs to the network consist only of the ten prior quarterly percentage changes in price and in earnings for each equity (by quarter, not accumulated), converted to a relative rank scaled around zero. Thirty simulated portfolios are constructed respectively of the 10, 20, …, and 100 top ranking equities (long portfolios), the 10, 20, …, 100 bottom ranking equities (short portfolios) and their hedged sets (long-short portfolios). In a 29-quarter simulation from the end of the third quarter of 1994 through the fourth quarter of 2001 that duplicates real-world trading of the same method employed during 2002, all portfolios are held fixed for one quarter. Results are compared to the S&P 500, the Value Line universe itself, trading the universe of equities using the proprietary "Value Line Ranking System" (to which this method is in some ways similar), and to a Martingale method of ranking the same equities. The cumulative returns generated by the network predictor significantly exceed those generated by the S&P 500, the overall universe, the Martingale and Value Line prediction methods and are not eroded by trading costs. The ANN shows significantly positive Jensen's alpha. All three active trading methods result in very high levels of volatility. But the network method exhibits a distinct kind of volatility: Though overall it does the best job of segregating equities in advance into those that will rise and those that will fall relative to one another, there are many quarters when it does not merely fail, but rather "inverts": It disproportionately predicts an inverse rank ordering and therefore generates unusually large losses in those quarters. The same phenomenon occurs, but to a greater degree, with the VL system itself and with a one-step Martingale predictor. An examination of the quarter to quarter performance of the actual and predicted rankings of the change in equity prices suggests while the network is capturing, after a delay, changes in the market sampled by the equities in the Value Line index (enough to generate substantial gains), it also fails in large measure to keep up with the fluctuating data, leading the predictor to be often "out of phase" with the market. A time series of its global performance thus shows antipersistence. However, its performance is significantly better than a simple one-step Martingale predictor, than the Value Line system itself and than a simple buy and hold strategy, even when transaction costs are accounted for.

# 1. Background

On a weekly basis, a wealth of technical and fundamental information on a representative universe of publicly-traded equities is updated each week, in principle, for every company in the well-known *Value Line Investment Survey* (VL, "the Survey") of approximately 1700 primarily-U.S. companies. According to a proprietary and not necessarily static formula known to depend disproportionately upon recent percentage changes in the price of an equity, the recent percentage change in its earnings, and especially on an intermittently-generated and more loosely quantified "earnings surprise factor"[1], the Survey updates and assigns to each equity every week a "Timeliness Rank" from 1 to 5 (In fact, not every equity is updated every week in consequence of a certain "slippage" in the VL system). This rank is a measure of future "price performance." Stocks assigned a rank of 1 are predicted to experience the largest positive *long*- and *intermediate*-term price change (six to twelve months), 5 the least (or greatest decline).

---

[1] jsatinov@princeton.edu
[2] dsornette@ethz.ch

Because the VL survey appears to provide information on equities with at least some predictive power, it has been the object of a significant amount of academic study, beginning with Shelton in 1969 [2], but most notably Fisher Black's 1973 paper, "Yes, Virginia, there is hope: Tests of the Value Line ranking system" [3] and a subsequent more detailed dissertation by a student of Black's at M.I.T. [4]. Other widely-cited studies have been performed again in 1973 (with a focus on risk [5]), and in 1981 (testing aggressive investing using VL ranks [6]), 1985 (testing the inverse effect: How VL rank changes affect stock prices [7]), 1987 (relating VL rank to firm size [8]),1990 (discussing the implications for the efficient market hypothesis—EMH), 1992 (relating the VL effect to post- announcement earnings changes [9]), 2000 (finding a positive effect even controlling for post-announcement earnings changes [10]) and 2008 (examining the predictive value of other data in the VL Survey apart from the ranking system proper)[11].

Until relatively recently, with the advent of more extensive computerized financial data services, the Value Line survey was one of the most widely-used for professional analysts' forecasts. It has been shown to provide some of the most accurate forecasts of analysts' predicted excess return, especially in comparison to other widely used sources (e.g., IBES, S&P) [12]. Fisher Black is reported to have offered the following advice in 1983: "One of the best ways for an investment firm to pilot a portfolio through the vicissitudes of the market would be to fire all the financial analysts, save one, and make that one read Value Line. [13]"

Even though VL defines its rankings to predict long-term price appreciation, most of these studies have concluded that its predictive power is real chiefly for the short-term only and only doubtfully so once transaction costs are included. Nonetheless, given the power of the efficient market hypothesis, the predictive capacity of the VL system is impressive: "In a world with no end of people hawking investment advice, the *Value Line Investment Survey* has captured the imagination of the finance community like few others[10]."

Studies of the VL ranking system consistently demonstrate that it is at least theoretically effective (before trading costs); and occasionally demonstrate that it is practically effective when its predictive range is carefully analyzed and applied such that trading costs do not erode gains[4].

Because the Survey claims to heavily weight "earnings surprises", and because these are known to affect prices, this factor has been offered as the explanation for how there could be some predictive power in the ranking system [14, 15]. But this explanation contrasts with VL's own arguments on its behalf, since the system claims to incorporate more than simply earnings surprises in arriving at its rankings (See for example, [1]). Indeed, most rank assignments are made without any earnings surprises. It appears that much of the outside research testing the VL ranking system on the basis of earnings' surprises presumes that the EMH is effectively correct—all *available* information is instantaneously incorporated into the present price of a stock; earlier price and earnings data therefore has no predictive power; an earnings surprise represents new (by definition *unavailable*) information that requires some time to be reflected in the current price; during this time the surprise therefore has (rapidly declining) short-term predictive power.

Nonetheless the question has been hotly debated in the academic community as to whether the VL ranking system as a whole can provide more than a theoretical refutation of the EMH. The semi-strong form of the EMH precludes the ability to profit from VL information as all such information—including VL's forecast of future price appreciation—would already be in principle incorporated into the present price of a security and thus discounted against future gains [16, 17]

A review of the literature makes it clear that while a significant majority of researchers have in fact detected a VL "anomaly" or "enigma", most find the size of the anomaly likely to be too small to be exploited given transaction costs [3, 4, 6, 10, 18, 19], regardless of whether it is attributed to the earnings' surprise factor or not.

The VL data and rankings are used both by analysts and traders: Given the relatively modest price of a subscription to the survey, VL could scarcely be a working business, especially for as long as it has been—since the 1960's at least—if its subscription base were only analysts). This leads naturally to another important consideration which has also been the object of study: the possibility of feedback between the weekly release of the VL rankings, especially therefore rank changes, and short-term price changes among equities undergoing rank changes. The EMH could remain in principle true, yet brief departures from it could occur simply because of the market response to rank changes—whether or not these changes accurately reflect underlying fundamentals.

Indeed, there is evidence that analysts will herd significantly (and thus their clients will trade accordingly) based on VL recommendations, thus amplifying any direct effect from VL subscribers. Among analysts who publish newsletters this herding unsurprisingly occurs when signal correlation among them is high. It also occurs, also unsurprisingly, when their measured performance ability is low. Perhaps more surprising is the fact that significant herding based on VL occurs when the analysts' *reputation* is high [20]. The surprise fades when one considers that high reputation may be as much an effect of herding, as independent of underlying fundamentals (ability) as may be prices.

In any event, there is evidence in the literature that a measurable component of the change in certain stock prices may be due a herding effect mediated by VL.

It should be noted, however, that the influence of VL may be declining. First, very little research on VL has taken place after the year 2000. Second, by 2000, popular discussion frequently noted a decline in the belief "on the street" that VL recommendations were still of use. In 2000, for example, *Money* magazine published an article decrying its decline[21]. The article made no academic claims, but may well have both represented and reinforced a popular belief that made its claims accurate. (In light of the above discussion, whether they were or not would depend of course in large part on how widely the belief remained that they were, or whether a consensus was arising that they weren't.)

In any case, the anomalous performance of the VL ranking system leads naturally to the question as to whether it may not be improved upon.

## 2 Review of the VL ranking system and its performance

The distribution of equities in each ranking "bin" is not flat. There are approximately 100 1's, 300 2's, 600 3's, 300 4's and 100 5's = 1500. The number of equities characterized as "3" varies the most, with other equities dropping in and out of the survey altogether over time. If the middle of the 3's is treated as a zero line, then the cumulative distribution of the ranks approximates a coarse-grained hyperbolic tangent: It is thus a natural way of quantizing the (discrete) rank-ordering by predicted price change for every stock, given the natural distribution of percentage price changes around zero, both positive and negative, ignoring the greater asymmetries at the extremes. (Since the smallest number of equities in the two 100-size bins are predicted to experience the largest price changes, respectively up and down, the next-sized bins the next largest price changes and the largest middle bin the least, the distribution of price changes—rather than bin sizes—conforms to a coarse-grained arc-tanh: If the largest positive changes are represented on the left side of a chart, the curve is actually a negative arc-tanh.)

During the period covered by this study (beginning of 1994 – end of 2001, i.e., capturing the run-up to, the peak and the drop-off following the 2000 "bubble"), VL rank 1 (T100, i.e., top 100), rank 5 (B100, i.e., bottom 100, Simulates long investments in Rank 5 stocks) or long-short (H100) hedged portfolios (long Rank 1, short Rank 5) would have performed as shown in **Figure 1**, assuming no transaction costs:

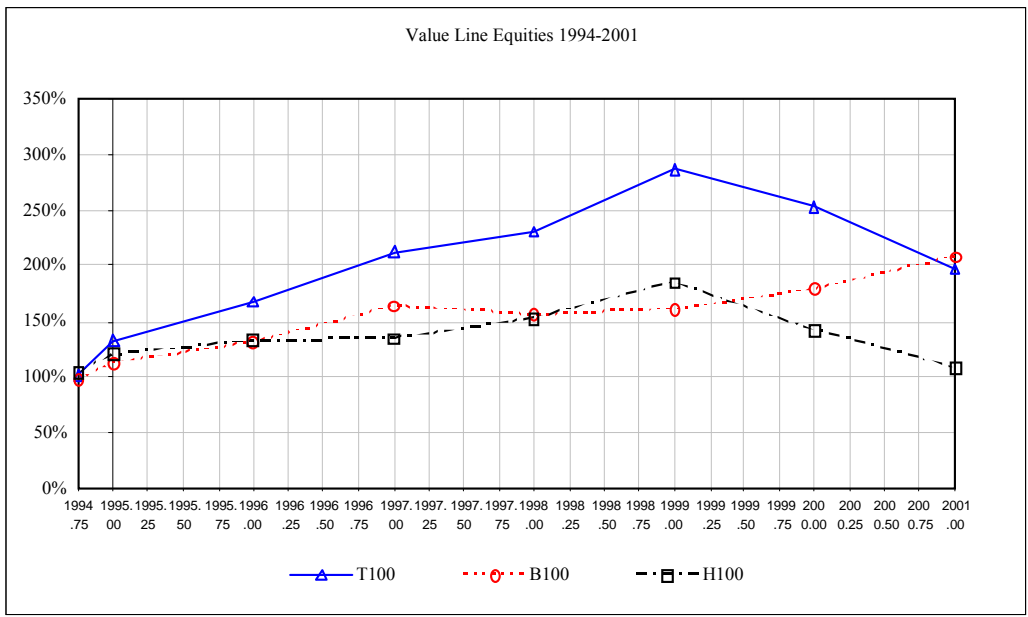

**Figure 1**: Returns from the VL Ranking System 1994-2001

During this period, reasonable costs would have eroded all competitive gains assuming weekly portfolio restructuring. Notice as well that while the slope of the compounded returns for the VL-defined T100 equities (rank 1) is > 1 for four-plus of the six-plus years, the slope of the compounded returns for the VL defined B100 equities (rank 5) is actually up for five+ years. More pertinent are the segments of relative slope, indicated qualitatively by the H100 return curve. This consists of five contiguous segments of positive slope followed by two of sharply negative slope where the ranking system not only fails, it inverts. As noticed by others and at other scales, inversions such as these are typical in the VL system ("In 1983 the average annual returns of stocks ranked four or five at the beginning of 1983 were higher than the corresponding average returns for stocks ranked one, two or three at the beginning of the year" [18]). The erosion of its competitive advantage is not primarily caused, for example, by intermittent "statistical" failures of the B100 portfolio to decline in absolute value and regress toward the market mean—especially easy to do in the face of a generally rising market. Rather, the failures and the erosion are abrupt and rapid—much faster than the gains as may be seen at a glance—and is caused rather by *reversals* in the model's performance. During these periods it seems, the model is not merely "not working", it is working "in reverse", as it were.

The VL investment survey includes many other kinds of rankings for its equities (e.g., "safety") and for broad categories of equities as well (e.g., performance and safety rankings for market "sectors"). Thus, another peculiarity of the VL approach to the financial domain—perhaps underestimated because of its simplicity—is that by combining and recasting so many numerical quantities into ranks, VL indeed performs a crude "renormalization" which—given the amount of noise, uncertainty and error in financial data—may nonetheless be quite effective, even if it was never conceptualized in such formal terms.

That is, VL's weekly "Timeliness Rank" (a function of relative change of price and relative change in earnings, inter alia) *might* be an effective method of weekly renormalization on a basis if it consisted of (or at least began with) a complete rank-ordering of the universe of equities, instead of a mere 5-bin coarse-grained version. The hand method of ranking originated by Arnold Bernhard has now been extended and computerized by his 86%-still-family-owned firm. The quasi non-linear, perhaps somewhat adaptive algorithms now employed remain proprietary. It is possible that the firm privately creates a complete and effective single-step ranking, but sells only the coarse-grained version to the public. If so, the remarkably poor performance of the VL *Fund* (a mutual fund that is said to employ the ranking system) argues against its efficacy. Or perhaps these proprietary methods involve the creation of the coarse-grained version only, or of something between it and a full rank-ordering. In any case, the coarse-grained version appears to contain enough information to be of academic interest and to provide a large incentive to subscribers, but to be at or just below the borderline of profitability once rigorously scrutinized, especially of late.

For example, we performed a preliminary Monte Carlo simulation on 1400 artificial stock prices undergoing randomized price changes drawn from the actual distribution of the Value Universe of price changes from 1994-2001, and then scattered into ranking partitions that mimic the 100-300-600-300-100 VL structure. This simulation shows that typically, 40-60% of the rank changes reported in the weekly VL survey can be attributed simply to price perturbations near the rank boundaries. Since on many weeks there may be only one or two changes, there are many weeks when no changes are caused by anything other than such perturbations. The Survey itself refers to this phenomenon somewhat misleadingly as the "dynamism" of the ranking system.

Of much greater interest—and pertinent to this study—is the fact that there is an immediate, very strong post-release effect on the price of a stock whose rank has changed once the change is announced. This effect has been noted and exploited by many (subscribers, analysts and their advisees)—and that can be exploited by VL company insiders completely legally in advance. It also has been examined and argued *not* to be caused primarily by a preceding earnings announcement: For instance, Thomas et al. [6] examined the impact of VL timeliness rank changes on stock prices while controlling for contemporaneous earnings releases, and found that the market response is consistent with increased liquidity in the shares.

Furthermore, our Monte Carlo simulation shows that stocks occasionally can change even two ranks stochastically because of the "tanh"-like distribution of rank bin sizes. Indeed, the distribution of the bin size ensures that these "meaningless" large events both happen and that they are disproportionately more likely to occur as moves both in and out of ranks 1 and 5—the ones that in turn have the largest post-release impact on the market they are meant to predict.

Thus, if there is indeed any genuine information contained in as coarse-grained a ranking system as that publicly available in the VL investment survey, it makes sense to attempt to create a more fine-grained version to extract it. The long history of the VL system with its tantalizing successes and failures; its longstanding and successful use of various ranking methods as naïve ways to handle scaling and normalization problems with financial data; the specific successes and kinds of failures of its "Timeliness Ranks" as a method for the short-term prediction of *relative* stock price changes; the observation of possible "phase" or "regime" changes causing catastrophic failure even in so crude a model as this; the tanh-like distribution of the bin sizes being suggestive of a coarse-grained renormalization of a finer-grained ranking; the fact that even though the VL system was developed nearly half a century ago, its core "inputs"—recent percent change in price and recent percent change in earnings—continue to dominate: All of these ideas suggest that it should be possible to create *de novo* a complete ranking of equities from the VL universe.

All of the above suggest that a very simple neural network architecture could be used to generate an equity ranking system that might provide insight into the phenomenon of *abrupt performance inversions* characteristic of the VL system and perhaps also improve upon the VL system itself by replacing its coarse-grained ranking with a more fine-grained version.

An initial system was developed for trading purposes and employed successfully as part of a number of hedge funds and funds of funds in different configurations in 2002. However, because of the frequent performance reversals at all time scales, in spite of continuing good returns, it was decided to perform a much longer and detailed set of studies of the method and the nature of the volatility of which this paper is part.

# 3 Methods

## 3.1 Equities

More than 1600 equities from the VL universe were selected with data eventually collected both by hand and from electronic sources from March 1, 1992 through December 1, 2001. All data was checked against two independent sources for consistency, and a third if discrepant (The primary source of data was the Value Line Investment Survey itself (the print version). The secondary source was Bloomberg, inc. Tertiary data sources were chiefly WRDS and Telescan). Equities with irreconcilable data discrepancy rates > 0.5% were eliminated from use. This resulted in a significantly smaller pool of equities than VL itself routinely uses in its ranking system and a much cleaner data set. From this universe, a permanently fixed set of 1452 equities were identified for data extraction.

However, in any given quarter, fewer than 1452 stocks may actually be ranked. This is always because of the listing of new corporations and the delisting of existing ones. To have included only equities that were listed throughout the test period (plus 10 prior training quarters for the first out-of-sample prediction = 39 quarters ≈ 10 years would have resulted in a very reduced set of equities highly biased toward large capitalization corporations unrepresentative of the VL universe.

## 3.2 Input data and outputs

Inputs to the network consist of the ten preceding quarterly percent price and earnings changes (not accumulated) transformed as ranks. Outputs are the predicted next quarter's percentage price changes. All ~1452 stocks are then ranked in descending order by the ANN's predicted percentage price change for the next (out of sample) quarter. (The MGL predictor simply uses the prior quarter's actual price-change rank as the best estimate for the next quarter.)

From this output, for each successive out of sample quarter, twenty portfolios are constructed (and ten more from hedged combinations among these twenty). The twenty portfolios represent cumulative deciles from 10 to 100 from the top and bottom ends of the ranking. A T10 portfolio consists of the 10 equities predicted to perform best, the T20 the twenty equities predicted to perform best, and so on to T100. The deciles are cumulative in the sense that the T20 portfolio consist of the T10 portfolio plus the next 10 best and so on. The B10,…,B100 portfolios are constructed similarly but from the bottom of the ranking up. H10,…H100 portfolios represented combinations of the respective T and B cumulative deciles with the T equities bought and the B equities sold short.

Depending on the week or month that the data is drawn from, raw price data will vary, of course, within a given quarter, whereas raw earnings data will either be unavailable, available, or will be available and then modified after the fact. Only original earnings reports were used, and only for those weeks and months in a cycle when they would actually have been available. Furthermore, based both on the well-known date of release problem in historical data, and on a variety of other glitches that arise in real-time trading (that were uncovered during experience with this method in 2002) approximately 30% of earnings reports that look as though they would have been available on a given day actually are not. Therefore, the final column of earnings data is not used at all as input for those weekly or monthly date cycles when it couldn't be available at all, and in all earnings input columns, 30% of the earnings figures are removed at random before ranking to simulate other real-life problems.

### *3.3 Selection of trading period start*

Given a starting quarter, there would in principle be (on average) twelve different weekly periods of data all starting in that quarter and sharing the same change in earnings value; or three monthly periods. (VL reports changes in its ranking system on a weekly basis.) The data structures for each of these cycles differ in their relation to earnings releases both with regard to the availability in relation to pricing data and from company to company. All of these considerations have been addressed, but because of the complexity of the task in back testing (by contrast to collecting data in real time going forward), the study reported here is limited to a single cycle of properly collected and error-checked data rather than an aggregation of between two and twelve weeks of data with an unknown amount of error and anachronism. The completed and fully error checked data set is simply the single best one able to be completed with the available resources. It contains no *known* errors. Back tests on other incomplete cycles show qualitatively similar results. The data period reported on here makes its first prediction for June 1, 1993 and its last for December 1, 2001 (roughly comparable to the report on hedge fund performance referenced below [22]).

### *3.4 Network Architecture*

The results reported on here are obtained using a simple back propagation network with a single hidden layer and recurrence. The results of multiple initializations are aggregated to obtain a final ranking. Exact net architecture and parameters are optimized independently on each new data set using a genetic algorithm but with extremely tight constraints. No variable deletion is allowed. Only a hidden single hidden layer is allowed. In general, minimal searching is permitted.

### *3.5 Training*

Training and testing sets are selected a-priori at random for optimization of training iterations. Under-fitting is greatly preferred to over fitting. Results are relatively insensitive to training lengths between six and twenty quarters. The results shown here are at about the median.

Two points should be emphasized here. First, as is an appropriate procedure in the use of ANNs, the network is always freshly trained on (ten quarters of) data that is out of the (one quarter) prediction sample. The ANN never has access to data from within the period it is predicting, hence the special care required with respect to using historical earnings data as explained in section 3.2.

Second, while the VL method as received by a subscriber appears static in the sense of implementing no evident adaptive or learning mechanisms, we know informally from a private meeting with the founder of the VL system that the regression-like formulae employed by VL are updated over time. (Thus any decline in its performance over time cannot be attributed to its algorithms become outdated solely because they are static).

# 4 Results

## *4.1 Hedged Returns*

### 4.1.1 Overall results

**Figure 2** provides a concise graphic snapshot of results, demonstrating the superior performance of the ANN predictor (not including transaction costs) relative to a one-step Martingale (MGL) predictor and (for all 100 equities) to the VL system itself, as well as to the S&P 500 index. Once a month, the ANN and MGL methods are used to predict and rank-order the top 10, top 20,…, top 100 equities (i.e., "cumulative decile": T10, T20, …, T100, from among the universe of 1452 stocks based on the inputs as described in section **3.2**), as well as the bottom deciles: B10, B20, …, B100. A portfolio is composed of matched T deciles held long and B deciles sold short (resulting in fully hedged "market-neutral" portfolios H10, H20, …, H100). Every portfolio is readjusted once per month. To adjust for possible monthly or seasonal effects, results are averaged over all three possible monthly starting points in a quarter. These results are compared to the actual VL selection of T100 and B100 stocks (groups "1" and "5" respectively) adjusted every quarter similarly (rotated and averaged), and to the S&P 500 Index over the entire out-of-sample range of 29 quarters. As shown here, trading costs are not included (to be discussed later). Note, however, that for small portfolios (i.e., H10, H20, H30), even were the turnover to be 100%, such costs are relatively moderate as they occur only at quarterly intervals.

**Figure 2** illustrates that at the end of the 29-quarter period, the ANN predictor succeeds at separating high-performing from low-performing stocks sufficiently well to generate substantial returns for all hedged decile ANN portfolios. A 6% annual risk-free rate of return has been assumed (high, therefore conservative). Furthermore, the internal progressively layered relations among the top 10, top 20, …, top 100 are very well-preserved by the network predictor: Returns generally fall off by cumulative decile (CD) implying that the separation is highly significant (These relations are not preserved by the MGL predictor which generates roughly the same loss for all cumulative deciles.

On the right edge, we see that the fully hedged T100+B100 = H100 equity portfolio using the network predictor yields annualized returns of 16.7% (10.7% in excess of the risk-free return). The MGL predictor for the same hedged portfolio yields excess annualized returns of 4.0% while the VL ranking system yields –0.8%

During this period both the broad market index (the S&P 500) and the (unweighted) VL universe performed roughly comparably, i.e., flat to slightly negative, thus we have eliminated any overall bias during the test period, but this result is composed of a period of rapid growth followed by a short period of high volatility followed by a period of rapid decline (before and after the 2000 market bubble), a challenging stretch of time for any model. The ANN not only does a superior job of ranking stocks than the VL method itself (while employing what is likely very similar inputs), in addition it parses the ranking more finely. (Both the ANN and MGL predictors provide explicit rankings for all 1452 stocks.)

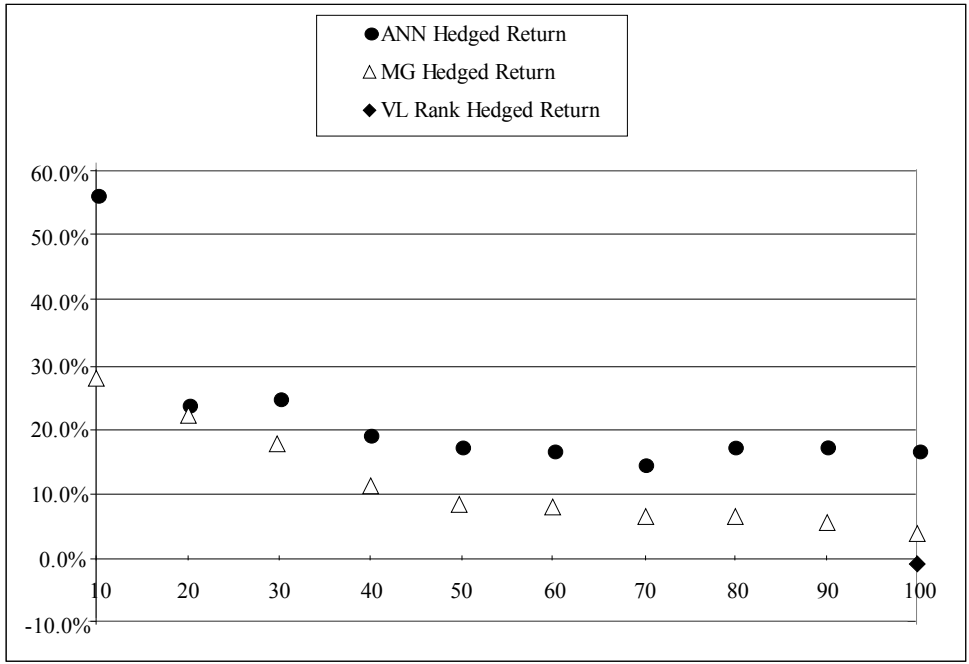

**Figure 2:** Comparative annualized excess returns from the ANN and MGL predictors for hedged (top and bottom 10, 20, …, 100) cumulative decile (CD) baskets of securities and for the VL system hedged Rank 1 and Rank 5 stocks.

To better understand abrupt performance reversals in the ANN and MGL, it will later prove instructive to decompose the hedge into its constituent long and short components.

### 4.1.2 Long Returns

**Figure 3** demonstrates the long-only ANN, MGL and VL system portfolios. Both the MGL and ANN predictors successfully generate the progressive relationship among cumulative deciles, but the ANN predictor generates superior returns in 9 of 10 instances. Comparing only the ANN Top 100, the MGL Top 100 and the VL Top 100, we find that the ANN returns 19.6%, the MGL 11.6% and the VL system 9.9% compared to the VL universe of 1452 stocks which returned –0.16% over this period. (The S&P 500, by comparison, returned –0.35%, i.e., both VL and SP are comparably flat). Thus, in a head-to-head comparison with the VL system (using all T100 stocks), the network predictor performs best by a large margin, the MGL predictor next best, the VL system comparably to the MGL and all three significantly better than the VL universe as a whole. Comparing the hedged to long only results, we see that the VL system has succeeded in identifying rising stocks but not in identifying poor-performing or declining ones. This fact is made evident by examining the short side of the results.

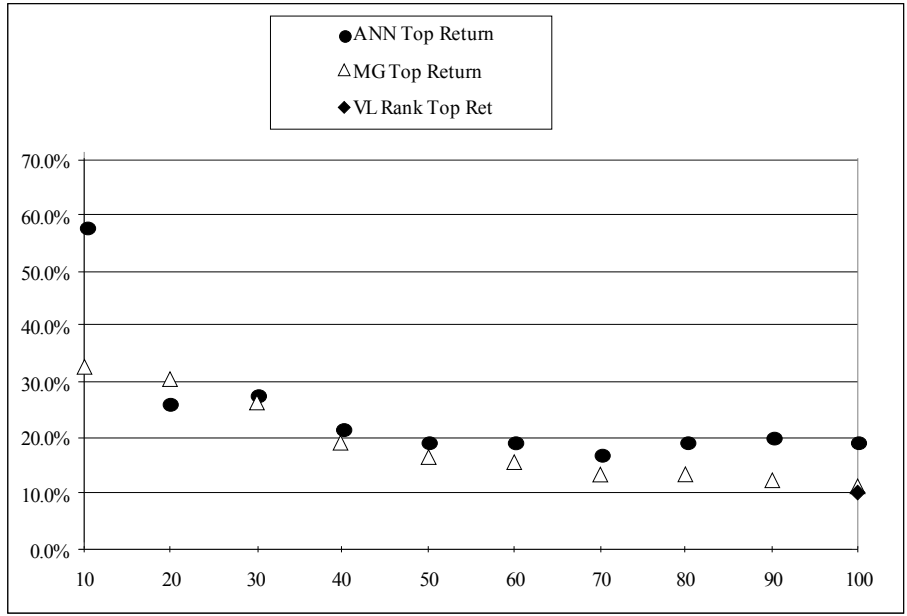

**Figure 3:** Comparative annualized excess returns from the ANN and MGL predictors for long (top 10, 20, …, 100) cumulative decile (CD) baskets of securities and for the VL system hedged Rank 1 and Rank 5 stocks.

### 4.1.3 Short Returns

In **Figure 4**. we illustrate the short-only ANN, MGL and VL system portfolios (results show the actual returns, not their negatives as is required for a short position). In other words, the lower the return the more desirable it is. We see that no system succeeds in generating absolutely negative results (it is typically far more difficult to predict stocks falling in price than rising ones). But the value of a hedged portfolio is not in amplifying gains by succeeding in shorting falling stocks. The goal is rather to create a portfolio that is "market-neutral" so as to neutralize price changes that may attributed to changes in the market as a whole. Captured gains therefore presumably arise from the intelligent selection of a portfolio of strategically chosen equities from among the available choices. (We will quantify the degree of success achieved by the ANN predictor in the next section.)

We see that the ANN successfully preserves the appropriate progressive relations among CDs: The bottom 10 are the worst performers (best for shorting), the bottom 100 the best (worst for shorting). The MGL predictor is not nearly so good as it was in identifying stocks for the long component—though it does a better job than the VL Rank 5 selection, its short baskets are all much better performing (therefore worse for shorting) than the ANN for all CD portfolios and the progressive relations among CDs are not preserved. Note, however, that the scale for the short selections is much compressed vis-à-vis the scale for long selection. The superior performance of the ANN versus the MGL predictor for hedged portfolios is therefore attributable both to its superior selection of top-performing sticks and bottom-performing stocks. The VL system fails altogether—its selection of bottom-performing stocks does better than the VL universe and even better than its selection of top-performing stocks.

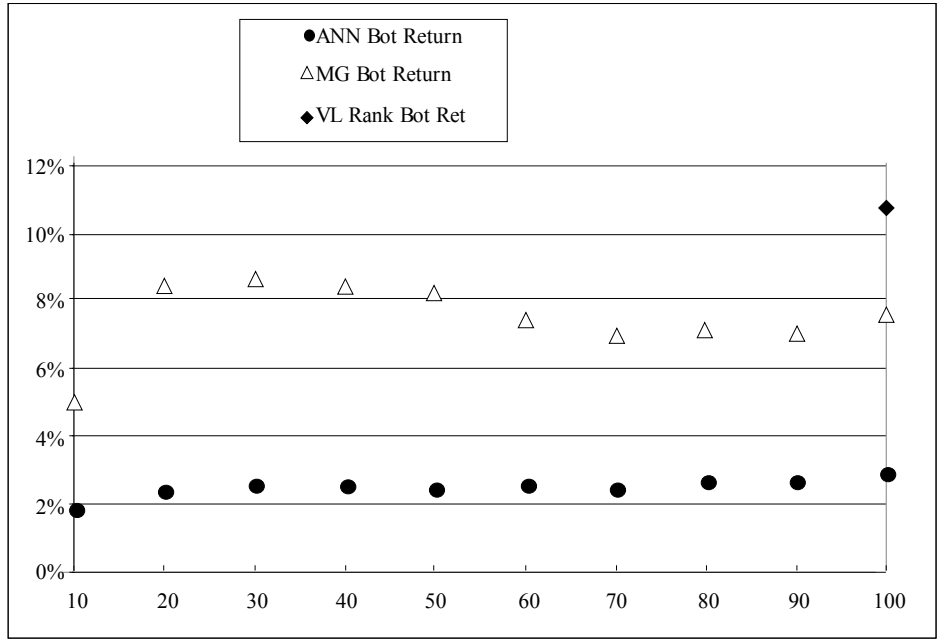

**Figure 4:** Comparative annualized excess returns from the ANN and MGL predictors for short (bottom 10, 20, …, 100) cumulative decile (CD) baskets of securities and for the VL system hedged Rank and Rank 5 stocks. Returns shown are as though long, not their negatives (shorted) as computed in the hedged portfolios. For the short component of the hedged portfolio, therefore, the lower the return the more desirable.

## *4.2 Risk-adjusted returns*

### 4.2.1 Sharpe ratios

The simple hedged returns by cumulative decile provide an excellent test of the capacity of the ANN to extract information about future stock performance from prior price and earnings changes. Note, however, that this model does not attempt to predict or to minimize risk, i.e., volatility. We do not necessarily expect that such a model will produce superior *risk-adjusted* returns. Indeed, a well-known correlation between performance and risk is the bête noir of most aggressive approaches to achieving superior returns. A perhaps overly simple but widely-used measure of risk-adjusted performance is the "Sharpe ratio", i.e., the ratio of the excess annualized return to the annualized standard deviation of the price. **Figure 5** provides a comparison of Sharpe ratios for the same categories as the excess returns in **Figure 2**. The general fall-off with cumulative decile is preserved once again for the ANN and more weakly for the MGL predictor. Except for the first cumulative decile, the resulting Sharpe ratios for the ANN predictor are lower than for the S&P 500 reflecting the relatively high volatility of the equities selected within the top 100. (Post-hoc examination reveals that these are very disproportionately in the technology sector which over this time period experienced unusually dramatic volatility in fact.) Thus, the superior returns generated by the ANN come at the cost of high volatility. (Note, however, that by this measure the ANN significantly outperforms the VL methodology for selecting high performance stocks as well as the MGL predictor and the VL universe as a whole.

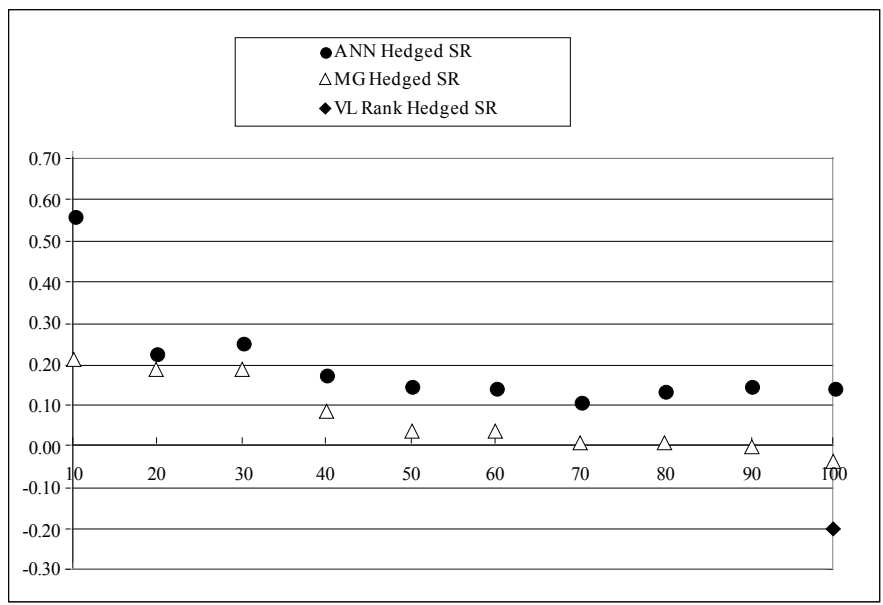

**Figure 5:** Comparative Sharpe ratios (risk-adjusted returns) for the ANN and MGL predictors for hedged baskets of securities; for the VL system proper; and for the VL and S&P 500 universes of equities. (VL H SR [green] and All VL SR [full] have the same value.)

Hedge funds in general claim to aim for and achieve Sharpe ratios of between 1 and 2. Again assuming a 6% risk-free rate of return, between January of 1990 and June of 2003, equity long/short funds reported annualized excess returns of ~12.07% [23] and an important study of eleven major market-neutral hedge funds from May, 1990 through April 2000 (hence during a period of generally rising markets only) reported Sharpe ratios of 1.1 [24]. By this measure, the ANN predictor falls short—unless one considers the very much more difficult and volatile period encompassed by the study. (It is very likely, furthermore that Sharpe ratios of 1-2 are less common than measured by various hedge fund tracking reports (for example, Hedge Fund Research) because data on failed funds is often unavailable, especially those that fail relatively quickly [23].

Another and arguably more accurate indicator of performance is provided by Jensen's alpha, and by the associated beta, a measure not of absolute volatility in terms of simple arithmetic or logarithmic price change, but of expected volatility given the volatility of the universe of stocks from which a portfolio is selected. From this perspective, the ANN succeeds remarkably well.

**4.2.2 Jensen's alpha**

The most widely used tool for assessing risk-adjusted investment performance is Jensen's alpha ($\alpha$). $\alpha$ is designed to quantify how an investment performs not absolutely but relative to the volatility of the actual market universe from which it is drawn. It has been widely demonstrated that high volatility investments with a large possibility of large losses are likely to demonstrate larger gains (upon success) than low volatility ones: The risk of losing a great deal is compensated for as a larger "risk premium". Thus, no investment "intelligence" is required to attain large gains by simply investing in a very high-risk vehicle. For example, a majority of start-up companies fail altogether. But those that succeed return a very high premium to their investors who assume the high risk. Well-established, so-called "blue chip" corporations offer relatively modest returns to purchasers of their stocks. But in exchange, shareholders may anticipate a relatively low risk of large losses.

The value added by a portfolio investment strategy (or manager) is therefore associated with gains beyond that attributable to the simple volatility of the appropriate universe of choices from which the strategy is drawn. $\alpha$ reflects the difference between the investment strategy's actual performance and the performance expected based simply on inherent risk. An investment that produces the expected return for the level of risk has an $\alpha$ of zero. $\alpha > 0$ implies that the strategy produced a return greater than expected for the risk taken. $\alpha < 0$ indicates that the strategy has produced a return smaller than expected relative to the risk.

It has been widely observed that ~75% of stock investment managers fail to improve on the performance of someone who had simply invested in a market-weighted basket of every stock. This phenomenon has been argued to be due to the "efficiency" of markets. This belief yielded market capitalization weighted index funds that seek to replicate broad market indices (i.e., baskets of securities representative of the entire pool of securities from which the selected ones are drawn), i.e., to reproduce the returns of those indices, hence aim for $\alpha = 0$.

$\alpha$ thus depends upon a measure of risk that is relative to a given market denoted Beta ($\beta$). The $\beta$ of an investment strategy is defined as:

$$\beta_{strat} = \frac{Cov[R_{strat} R_{mkt}]}{Var[R_{mkt}]} \qquad (1)$$

where $R_{strat}$ is the return of the strategy and $R_{mkt}$ is the return of the market from which the strategy is drawn. In other words, $\beta_{strat}$ is the slope of the linear fit of a scatter-plot with $R_{strat}$ the abscissa, $R_{mkt}$ the ordinate.

Stock index funds established a tacit standard of performance for investment strategies (and their managers): The successful strategy is one that yields performance in excess of the passive strategy of investing in everything equally since the latter strategy is more likely to be the better one, statistically. $\alpha$ is thus any additional return above the expected return of the $\beta$-adjusted return of the appropriate market.

The formal expression for investment $\alpha$ is derived from the Capital Asset Pricing Model (CAPM), wherein the estimated return $R_i$ on a security $s_i$ is given by three terms:

$$R_i = \alpha_i + \beta_i R_{mkt} + \varepsilon_i \qquad (2)$$

$\alpha_i$ (in CAPM) is considered a constant drift unique to the $i^{th}$ asset in the market with asset weights $mkt \equiv \{s_1, s_2, ..., s_i, ... s_N\}$; $\beta_i$ is the volatility of asset $i$ as defined in (1); $\varepsilon_i$ is a random variation unique to asset $i$ with mean zero (hence as the number of different assets increases in a portfolio of many securities, the combined $\varepsilon_i$ for the portfolio vanishes). Thus:

$$\alpha_{strat} \equiv R_{strat} - \left[ R_{rf} + \beta_{strat} \left( R_{mkt} - R_{rf} \right) \right] \tag{3}$$

where returns have been adjusted by the risk-free rate of return, $R_{rf}$. By this standard, $\alpha_{strat} > 0$ represents a successful strategy.

The $\alpha$ results for the ANN predictor are especially illuminating, given that the study period was one of unusual volatility. These are presented in **Figure 6**. The results show $0.4 > \alpha_{ANN} > 0.2$, roughly half the value reported by eleven major market-neutral funds between the bull-market of May, 1990 and April, 2000 ($\alpha \approx 0.6$), but still significantly positive (and in line with the Sharpe values presented above). More importantly the preservation of progressive relations among CDs is preserved more rigorously and for this measure of risk-adjusted performance, a power-law fall-off has a high correlation coefficient. (The market for which we calculate $\beta_{ANN}$ to derive $\alpha_{ANN}$ is the VL universe.) As we may anticipate from **Figure 5**, $\alpha$ for the MGL predictor is poor, hovering at or below zero throughout, a reflection of the excess volatility associated with its returns. The fit shown in **Figure 6** as the dotted line reads $\alpha = 0.35 \times CD^{-0.25}$ with $r^2 = 0.94$.

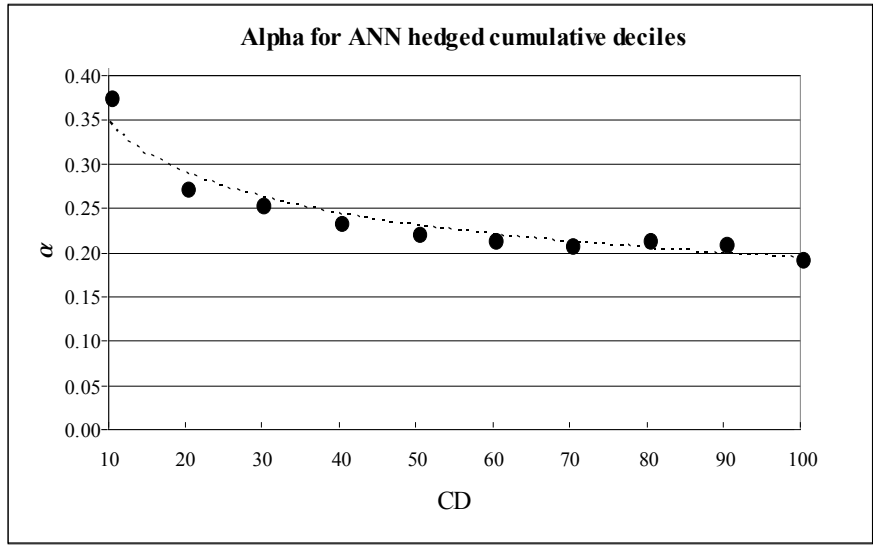

**Figure 6**: $\alpha$ as a function of CDs for the ANN with $\beta$ for the VL 1452. (MGL results not shown as they are evidently worse).

We now turn to the question of abrupt reversals in predictor performance.

## 5 Antipersistence in predictor behavior

This study was initially motivated by two facts: First, that a subtle feature of the VL system is that even if it has a history of in general working (i.e., making successful predictions contrary to the EMH), when it fails its failures are especially striking. Second, that the ANN predictor described above, which created a more fine-grained version of the VL rankings, and which was traded successfully in the real-world (with net positive post-transaction-cost gains) likewise suffered from high volatility—i.e., its periods of success were remarkably large but were interspersed with periods of large failure.

In section 4., studying a much longer period of time, we have shown that in spite of high volatility, the ANN predictor nonetheless is able to generate significantly positive $\alpha$. Thus, its gains, and implied predictive capacity, can not be attributed solely to the volatility ($\beta$) of the underlying market in which it trades.

We wish therefore to understand what is the nature of the swings in the returns generated by the predictor. To this we end, we first examine in finer detail the structure of the rankings generated by the ANN across all ~1500 equities. We argue that the peculiar distribution of actual returns by predicted rank may be understood *within each time period* (quarter) as a weighted mixture of highly successful and highly unsuccessful predictions. Whether a time-period's out-of-sample prediction is on balance successful or not depends upon the relative weights of the successful and unsuccessful components of the ranking. The respective components may be represented mathematically as negative and positive ArcTanh functions.

We then examine the decomposition of the rankings across time and show that the balance between successful (net negative ArcTanh distribution of returns by ranking) and unsuccessful (net positive ArcTanh distribution of returns), varies to a degree that is greater than what should be expected by chance, in particular showing "anti-persistence". Persistence is defined as a measure of the tendency of a binary series to exhibit repeating patterns; anti-persistence the tendency to specifically avoid repeating patterns. A random sequence exhibits neither persistence nor anti-persistence.

We then concentrate on just the extreme left (T10-T100 portfolios) and right (B10-B100 portfolios) ends of the distributions, combining the data across time into two sets: one set representing all the net-successful quarters, a second set representing all the net unsuccessful quarters. From the hypothetical returns generated by these two contrasting sets, we find that the nature of the failure of the ANN predictor, when it fails, is different from and less dramatic than the more complete inversion of results of the MGL predictor, especially in choosing equities for the B10-B100 portfolios.

## *5.1 Distribution of ANN predictions as a composite of success and failure*

For each of the 29 out-of-sample quarter, both the MGL and ANN predictors forecast the quarterly price change for every stock. The stocks are then sorted in descending order by this forecast to obtain the predicted rank-ordering. **Figure 7** shows the distribution of actual returns for all equities by ANN-predicted rank. The left plot shows the mean results for all quarters averaged over each rank; the right plot shows quarter 26 as an typical example for a given quarter. Within both plots those with the highest predicted rank are to the left, those with the lowest to the right. The (varying) butterfly-shaped distributions are typical for both the MGL predictor and the ANN predictor. The precise shape and density of the distribution determines whether the predictor is successful for long, short or hedged portfolios and for which CDs. Rising on the left and declining on the right is most desirable (successful prediction of both rising and declining equities); a greater rise on the left than on the right represents the overall results where the selection of rising equities succeeds to a greater degree than the selection of declining equities fails, enough to yield positive results for a hedged portfolio.

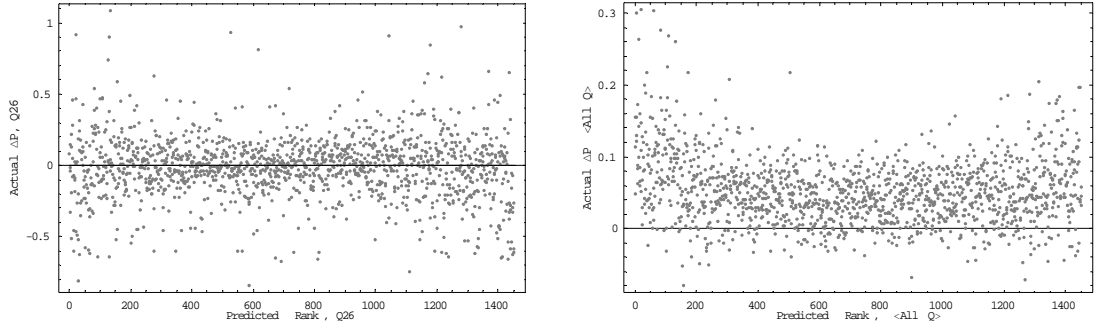

**Figure 7:** All 1452 actual returns (vertical axis) by ANN-predicted rank (horizontal axis) in quarter 26 (left) and for all quarters (right).

The butterfly shape creates a difficulty in fitting the distribution meaningfully to a single function. We propose that this distribution results from a "superposition" of successful and failed predictions with the predictor being simultaneously both unusually successful (better than chance) and unusually unsuccessful (worse than chance). The large degree of scatter reflects the closeness of the balance between the two and reflects the ease with which global success tips over into global failure and vice-versa leading to frequent reversals.

Suppose that the predictor were 100% successful: i.e., the rank-ordering predicted by the ANN corresponded exactly to the descending ordering of actual price-changes. In that case, the points in **Figure 7,** instead of forming a scattered distribution, would create a monotonically descending curve from left to right. On the other hand, were the predictor 100% unsuccessful, the points would form a monotonically ascending curve from left to right. These perfect functions (for the mean by rank over all quarters) are shown in **Figure 8**.

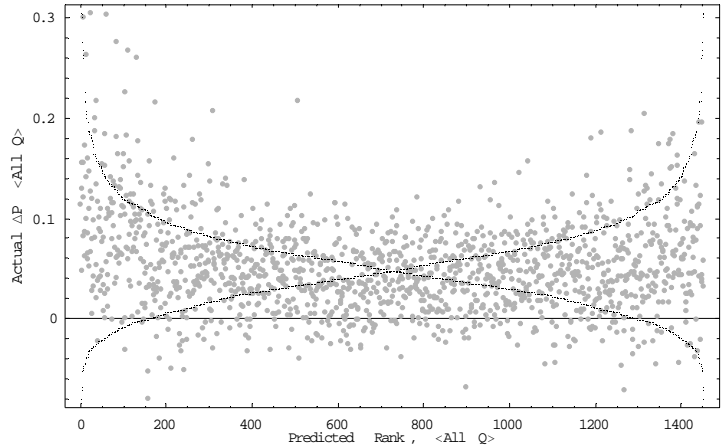

**Figure 8:** All equities for quarter 26 in descending (exactly correct) order of actual price change and ascending (exactly incorrect) order superimposed on the actual distribution.

The actual butterfly distribution then may be considered a weighted mixture of the fully correct and fully incorrect distributions. (Define the weight of the fully correct distribution as $\theta_1$ and of the fully incorrect as $\theta_2$.) To find $\{\theta_1, \theta_2\}$ empirically, we first fit both the correct and incorrect distributions to respective ArcTanh functions with parameters that thus reflect the actual distribution for the quarter. We find that:

$$\Delta p_{correct} = -0.254 + 0.346 \times ArcTanh(1 - 0.000689 r_{correct}) \quad (4)$$

where $\Delta p_{correct}$ is the actual price change as predicted by the correct ranking and $r_{correct}$ is the correctly predicted rank.

Likewise:

$$\Delta p_{incorrect} = 0.227 - 0.350 \times ArcTanh(1 - 0.00689 r_{incorrect}) \quad (5)$$

with $r_{incorrect}$ the rank predicted maximally incorrectly. (In both (4) and (5) the subscript on $\Delta p$ refers to the source of the price change, either the correct or incorrect ranking. In both cases the $\Delta p$ in question is one of the actual ones.)

We may now fit the actual distribution of price changes by the actual ANN prediction/ranking to a weighted sum of these two functions (for which the ArcTanh terms are of course the same):

$$\Delta p = \theta_1 \Delta p_{correct} + \theta_2 \Delta p_{incorrect} = 0.575 \Delta p_{correct} + 0.425 \Delta p_{incorrect} \quad (6)$$

which yields the curve shown in **Figure 9**. (The same fit may be obtained directly from a single ArcTanh fit, but this would not yield the weights of the decomposed correct and incorrect components.)

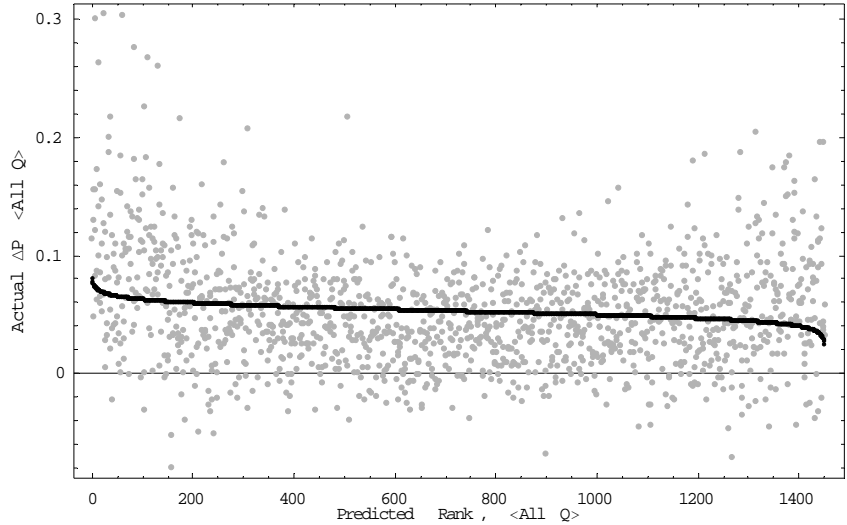

**Figure 9:** ArcTanh fit to the actual price change by ANN predicted rank for the out-of-sample data in quarter 26.

In greater detail, we find that $\theta_1 = 0.575 \pm 0.018$, $\theta_2 = 0.425 \pm 0.018$, $r^2 = 0.071$, $CI(\%, \theta_1, \theta_2) = \{95, 0.54 < \theta_1 < 0.61, 0.39 < \theta_2 < 0.46\}$ (so that in spite of the scatter the weighting remains non-overlapping at the 95% confidence level), $\chi^2 / DOF = 0.002$ with a weakly significant $p < 0.05$. These statistics are all superior to a simple linear fit, the slope of which nonetheless also quantifies the general principle that points to the left should on average be above those to the right if the predictor is working. The central part of the distribution is in fact essentially linear and close to zero; decreasing/increasing tangents to the fit-curve at the respective left and right extremes better represent those equities that constitute the T, B and H 10-100 portfolios we study in detail.

The fact that the fit curve is generally declining from left to right implies that over the entire list of equities the predictor is on average successful. The fact that the curve is almost entirely above the zero line illustrates that the universe of equities gained overall in value during the study period.

Each quarter's rank ordering constructed by the ANN may be similarly treated as a mixture or "superposition" of a globally successful and a globally unsuccessful distribution. The net balance of each yields both the distinctive final result and generates the typical butterfly distribution.

The question may be raised as to whether such a fit is practically meaningful. The most straightforward evidence that it is (averaged across all quarters)—and that it is also both meaningful in the MGL equivalent and in comparison shows evidence of the MGL predictor's poorer performance—follows from the three facts that (1) the ANN generates significantly positive returns for the hedged 10, 20,…100 portfolios (left and right ends); that (2) the same is true for the MGL predictor, but less so; and (3) that the returns among the 10 CDs for both the ANN and MGL generally fall off by CD. Furthermore, we discuss in section 5.4 similar graphed results examining *only* the top and bottom extremes of the distribution but aggregated for all quarters, segregated by winning and losing quarters. A negative ArcTanh-like distribution is visibly evident for the aggregated winning quarters likewise a positive Arc-Tanh-like distribution for aggregated losing ones. Comparing these similar results between the ANN and MGL we find in the different structure of the distribution an explanation for how the ANN outperforms the MGL.

## *5.2 Anti-persistence in predictor performance*

The weights for the successful and unsuccessful components in each quarter's prediction always add to 1. We may therefore track the weight of the successful component to glean a snapshot of ANN performance over time as in **Figure 10**.

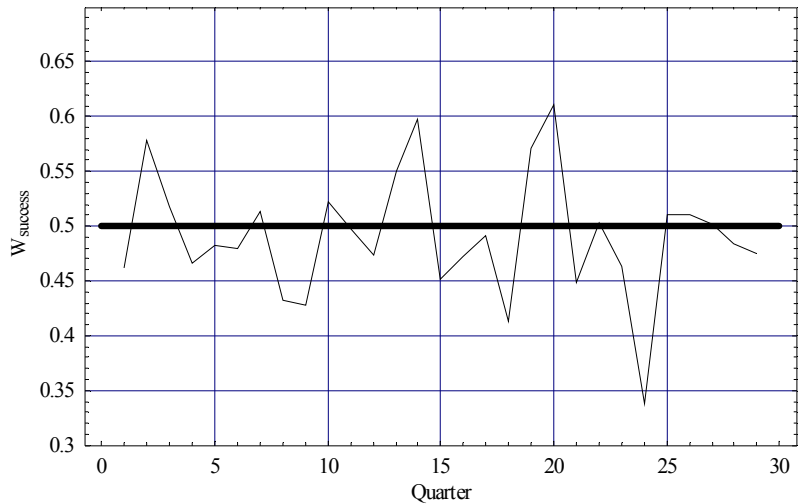

**Figure 10:** Proportion of successful component ($W_{success}$) of ANN predictor by quarter.

Impressionistically, we see that the predictor appears to show relatively wide and frequent swings between being globally successful and not. We may quantify this impression by considering the sequence of "successful" ($W_{success} > 0$) and "unsuccessful" ($W_{success} < 0$) quarters as a binary series and examining this series for persistence.

As discussed in [25-27], persistence is a formally defined measure on [0,1] of the extent to which patterns in a binary series tend to repeat, anti-persistence of their tendency to avoid repetition. The persistence $\mathfrak{P}$ of a maximally repetitive sequence equals 0, of a maximally anti-persistent sequence 1. A random sequence has persistence $\mathfrak{P} = 0.5$.

In brief, persistence is determined at a particular scale $m_s$ by examining all possible binary $m_s$-bit subsequences and counting the proportion of times that when a particular subsequence is followed in the series by 0, it is followed by 0 again upon its next occurrence; likewise for a following 1. Details of the calculation and examples from other domains may be found in [25-27]. At a scale of 1 (which is the only scale at which so short a sequence can have a meaningful persistence measure), $\mathfrak{P}$ measures the tendency of a series not to alternate ($1 - \mathfrak{P}$, its anti-persistence, the tendency to alternate, with the series 1,0,1,0,1,0,… or 0,1,0,1,0,1,… having $\mathfrak{P} = 0.5$).

We find that the 29 quarters of predictor data in **Figure 10** have $\mathfrak{P}_{MGL} = 0.143$ at $m_s = 1$ which is highly anti-persistent. It might appear that with only 29 binary values, the measured $\mathfrak{P}$ could not be statistically significant. But in fact, measuring $\mathfrak{P}$ on 1,000,000 random binary sequences each of length 29 yields $p \leq 0.004$. A similar set of analyses performed on the results from the MGL predictor yields $\mathfrak{P}_{MGL} = 0.357$ with $p \leq 0.0013$.

The implication of a significant degree of antipersistence in a predictor's results is that whether returns are in general positive or not, they are associated with frequent performance reversals, a particular kind of volatility that is highly undesirable. The MGL predictor's higher volatility as measured by its Sharpe ratio is consistent with its somewhat greater anti-persistence at scale 1. (Note that anti-persistence in general, i.e., at larger and at many different scales, tends to be associated with lower volatility. Here, we are concerned with the phenomenon of abrupt performance reversals from quarter to quarter.) We see in the comparison between anti-persistence in the ANN and in the MGL the fact that while high performance tends to be associated with frequent performance reversals, it is possible for subtler methods to yield a more satisfactory relationship between volatility and return as reflected in the measure of alpha.

A natural question is whether there is correlation between successful versus unsuccessful predictions by quarter and the direction of the market. Perhaps the ANN is successful when the market rises and unsuccessful when it falls—a common complication of naïve predictors. If the direction of the overall mean VL universe by quarter—equivalent to a buy and hold strategy—is converted into a binary series, this series is also anti-persistent with $\mathfrak{P}_{All} = 0.283$. For a series of this length, this degree of antipersistence is not highly statistically significant $p \leq 0.035$, suggesting that the variation in quarterly mean returns may well be effectively random. Furthermore there is no correlation between success by quarter for the ANN and mean gain by quarter: $r^2 < 0.00001$.

## 5.3 Effects of overall market performance on the top and bottom ends of the predictor rankings

The severity of the models failures when they fail (both the ANN predictor and the MGL) reflects the fact that the state of the model is at times of failure at best one time-step behind the phase of the market. Nonetheless, if its inputs have been properly chosen so as to reflect actual feedback characteristics of the market in question such that by tacitly learning patterns of market change (which a MGL predictor cannot do), the ANN model should be able to change its state rapidly enough to compensate for those time periods when its state is out of phase with the market and generate a net cumulative positive return that is significantly greater than any control, as has occurred.

We know that at least some real-world feedback has been demonstrated between the selection of top-ranked VL stocks and changes in its price and earnings reports. The ANN inputs have been chosen with this in mind and look backward over ten quarters' worth of prior earnings and price rankings, developing a tacit relation among these in determining its prediction for the subsequent ranking. Unlike the MGL predictor, it may detect a pattern not only of ranking but of change of rank structure over time. (We do not here report on a similar predictor that does not use earnings data. Results are degraded in the direction of the MGL predictor but remain superior to it.)

In order to quantify results efficiently, and test this hypothesis using so limited a number of quarters, the quarterly results can be segregated by positive or negative net return for the market as a whole (the VL universe) and then aggregated. If there is a strong tendency for quarters with net positive returns to show a negative ArcTanh-like distribution (and an positive ArcTanh for quarters with negative returns), aggregation should amplify this effect. Since quarters with results close to zero are included, considerable noise is introduced, strengthening the significance of the aggregation while ensuring that no selection bias is involved in the segregation. We do not attempt to fit our results, since as will be demonstrated, we are combing data from the extremes of the distribution, creating a discontinuity.

Aggregated results by top, bottom and hedged 10 through 100 deciles can be compared to the MGL predictor for the same groupings likewise segregated into the same two major classes (rising quarters versus declining quarters).

In preparation for this analysis, we first provide some relevant global measures of performance.

**Figure 11** illustrates the cumulative returns over time for the hedged ANN portfolios (log scale). Note the tendency for the returns to be relatively extreme in both directions; for quarters with sharply positive (negative) returns to be followed by quarters with sharply negative (positive) returns; for the progressive relationship by decile to be relatively well-preserved through time; and for the net cumulative return in all fully balanced long-short portfolios to be significantly positive. Note too how the progressive relations among cumulative deciles are quite well preserved over time.

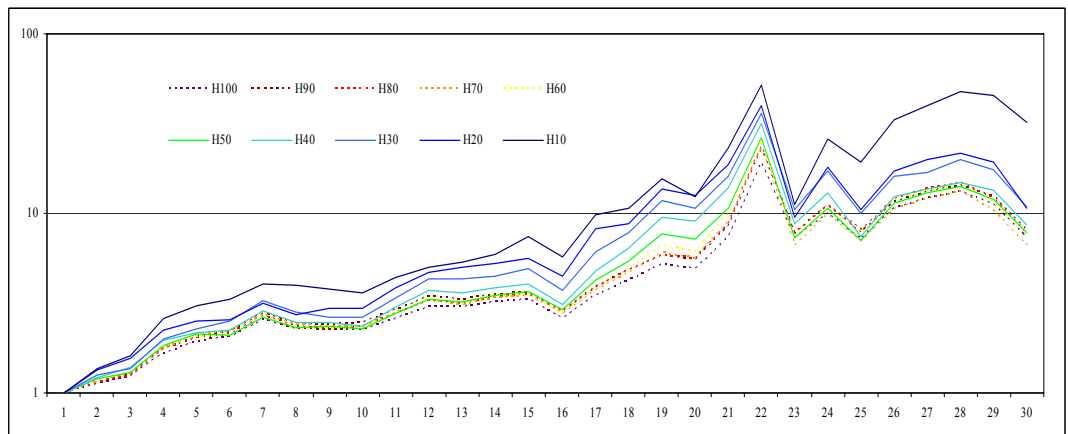

**Figure 11:** Semi-log Chart of Multiple of Initial Investment in Hedged Portfolios over 29 Quarters (color online; above chart shows general preservation of CDs—note poorer performance of dashed versus solid return curves.)

**Table 1** shows the cumulative decile performance for the ANN and MGL predictors, followed by their differences. We will want to analyze what contributes to this difference in performance. We will see that the two predictors have many similarities that contribute to both their successes and their failures, but in varying proportion. In quantifying these proportions we can gain some insight into the phenomenon of change of state.

**Table 1**

| Portfolio | Excess Ann'l Return | Portfolio | Excess Ann'l Return | Difference |
|---|---|---|---|---|
| ANN H 10 | 50.2% | MGL H 10 | -10.2% | 60.4% |
| ANN H 20 | 18.1% | MGL H 20 | -10.0% | 28.1% |
| ANN H 30 | 19.2% | MGL H 30 | -9.6% | 28.8% |
| ANN H 40 | 13.2% | MGL H 40 | -10.4% | 23.5% |
| ANN H 50 | 11.3% | MGL H 50 | -10.6% | 21.9% |
| ANN H 60 | 10.9% | MGL H 60 | -10.6% | 21.5% |
| ANN H 70 | 8.7% | MGL H 70 | -10.8% | 19.5% |
| ANN H 80 | 11.0% | MGL H 80 | -10.7% | 21.8% |
| ANN H 90 | 11.5% | MGL H 90 | -10.8% | 22.3% |
| ANN H 100 | 10.7% | MGL H 100 | -11.1% | 21.8% |

As noted before, the MGL method does reasonably well in selecting equities that simply *continue to rise*—especially as part of a universe of equities experiencing a general rise. But the method does not appear capable of differentiating those that will significantly rise (relatively) from those that will significantly fall, a weakness that it shares with the VL predictor than whose final results it does scarcely any better. We will see that the ANN shares this same weakness, but in lesser degree. This particular kind of failure appears in two guises:

1. The failure appears cross-sectionally within any given prediction set in the form of a series of equities predicted to make large price changes in one direction that actually make large price changes in the opposite direction (giving rise to the anti-persistent behavior discussed above). Typically, such mistaken predictions are admixed with a large number of correct predictions. This gives to the overall scatter plot of rank-orderings versus actual percentage price changes a butterfly shape from which quantifiable information will be extracted for the ANN predictor, the MGL control and their differences. The weighted proportion of successes and failures in the top 10, 20,…,100 portfolios versus the matching bottom portfolios determines the magnitude of the success or failure of the respective hedged outcome for any given quarter.

2. The failure appears intermittently and abruptly through time as the balance between overall correct and incorrect predictions shifts. As we will see, this shifting balance produces large changes in outcome from quarter to quarter generating the impression of a change in state that is especially pronounced in the MGL model. The final results are a consequence of the accumulated successes and failures over time.

## 5.4 T/B Portfolios aggregated by success or failure of the predictor(s)

### 5.4.1 T/B 10 = H10 Portfolios

One gains an intuitive impression that the universe of equities tracked by the model(s) undergoes sudden changes in performance partially captured in the "whipsaw" behavior of the predictors generated by the models. The challenge is to devise and extract a simple measure that objectively characterizes this phenomenon, if present.

We have taken the following approach: All ~1500 equities have been assigned out-of-sample predicted ranks for all 29 quarters both by the ANN and the MGL predictors. All equities likewise have their known in-sample actual percentage price-changes for every quarter. We thus start with two ~1500 X 29 tables showing the rank (by row number, with the best predicted rank being 1, the worst ~1500), one table for the ANN and the other for the MGL. From these tables we keep only the top 100 rows and the bottom 100 rows. We create an identical set of tables this time keeping only the top 90 rows and the bottom 90 rows. Again for the top and bottom 80, 70, …, 10 rows. Within each table, we identify those quarters where counting all ~1500 equities, the mean market change is positive and where it is negative.

Table 2 shows a hypothetical example for the T100 and B100 price changes sorted by predicted rank: Gray columns represent quarters where the mean price change for the portfolio—long on the top 100, short on the bottom 100—is negative. White columns represent positive quarters.

**Table 2**: A data table with (hypothetical) actual price changes arranged by predicted rank and quarter for top and bottom 100 equities.

| ↓ Rank/Quarter ⟶ | Q1 | Q2 | Q3 | … | Q29 |
|---|---|---|---|---|---|
| 1 | +.001 | +.002 | −.004 | … | −.014 |
| 2 | −.017 | +.018 | −.050 | … | +.005 |
| … | … | … | … | … | ... |
| 100 | −.003 | −.009 | −.011 | … | −.014 |
| 1353 | +.020 | +.017 | +.021 | … | −.014 |
| … | … | … | … | … | −.003 |
| **1451** | +.008 | +.015 | +.034 | … | +.001 |
| **1452** | +.045 | −.003 | −.016 | … | +.022 |
| **Mean Hedged Ret.** | −.043 | +.032 | −.004 | … | +.007 |

For each data table (2 tables for each of the size 100, 90,…,10 data sets), we then segregate the successful and unsuccessful quarters (+ or − net or mean gain for the hedged portfolio) as in **Table 3** and **Table 4**.

**Table 3:** Successful quarters only

| ↓ Rank/Quarter ⟶ | Q2 | … | Q29 |
|---|---|---|---|
| 1 | +.002 | … | −.014 |
| 2 | +.018 | … | +.005 |
| … | … | … | … |
| 100 | −.009 | … | −.014 |
| 1353 | +.017 | … | −.014 |
| … | … | … | −.003 |
| 1451 | +.015 | … | +.001 |
| 1452 | −.003 | … | +.022 |
| **Mean Hedged Ret.** | **+.032** | **…** | **+.007** |

**Table 4:** Unsuccessful quarters only

| ↓ Rank/Quarter ⟶ | Q1 | Q3 | … |
|---|---|---|---|
| 1 | +.001 | −.004 | … |
| 2 | −.017 | −.050 | … |
| … | … | … | … |
| 100 | −.003 | −.011 | … |
| 1353 | +.020 | +.021 | … |
| … | … | … | … |
| 1451 | +.008 | +.034 | … |
| 1452 | +.045 | −.016 | … |
| **Mean Hedged Ret.** | **−.043** | **−.004** | **…** |

Within each portfolio H10 – H100, but now segregated into all successful quarters and all unsuccessful quarters, we examine the top component of the portfolio and the bottom component (and do so for both the ANN and MGL predictors). We quantify the performance *across* all successful quarters of the rank 1 stock, the rank 2 stock,….through the rank 10, 20,…,100 stock (depending on the size of the portfolio we are studying). Then we do the same with the bottom groups. We likewise quantify the performance in the same way across all unsuccessful quarters. This will allow us to examine the structure of the each predictors' assignment of rank not quarter-by-quarter, but treating successful and unsuccessful quarters (as we did in sections 5.2 and 5.) but in the aggregate instead, and each aggregation separately.

The simplest way to thus aggregate the data across rows is to use their mean and standard deviation (which we will examine shortly). But this would fail to account for the difference in the number of winning and losing quarters (and the difference in these numbers between the ANN and MGL predictors) which after all is a key consideration. The measure employed here is therefore an artificial metric adapted from the "natural" features of the financial domain: Pseudo-compounding of the numbers across a row (plus one each) as though all winning quarters occurred in sequence following an initial investment of 1; all losing quarters similarly. This method is meant to place all quarters on an independent footing while properly weighting the geometric effect of an imbalance in the number of winning and losing quarters. The resulting numbers will therefore scale somewhat like the actual results but are not "real": Both the winning and the losing quarters, for both MGL and ANN tables have simply been compounded in sequence from the actual percent changes. Unrealistically sized final gains and losses for each row are then used in place of a mean, but the relative relations among points are properly preserved. while being pushed apart around 1.

**Figure 12** shows the results of the MGL predictor for the Top and Bottom 10 portfolios under this transformation. Each mark represents a pseudo-compounded return for a quarter with full marks for the aggregated successful quarters ("+"); hollow marks for the unsuccessful ones ("−"). Keeping in mind the discussion in sections 5.1 and 5.2, we see something similar: The full marks all together (from both top and bottom ends of the ranking, joined together at the midline, as it were) create a negative ArcTanh-like distribution, the hollow ones a positive ArcTanh. The full mark distribution thus mimics a successful quarter when the predictor is "in phase" so to speak with the distribution of price changes; the hollow mark distribution mimics an unsuccessful quarter when the predictor is "out of phase" with the distribution of price changes. But in this case we have aggregated the data from all quarters, collating the "in phase" and "out of phase" results separately.

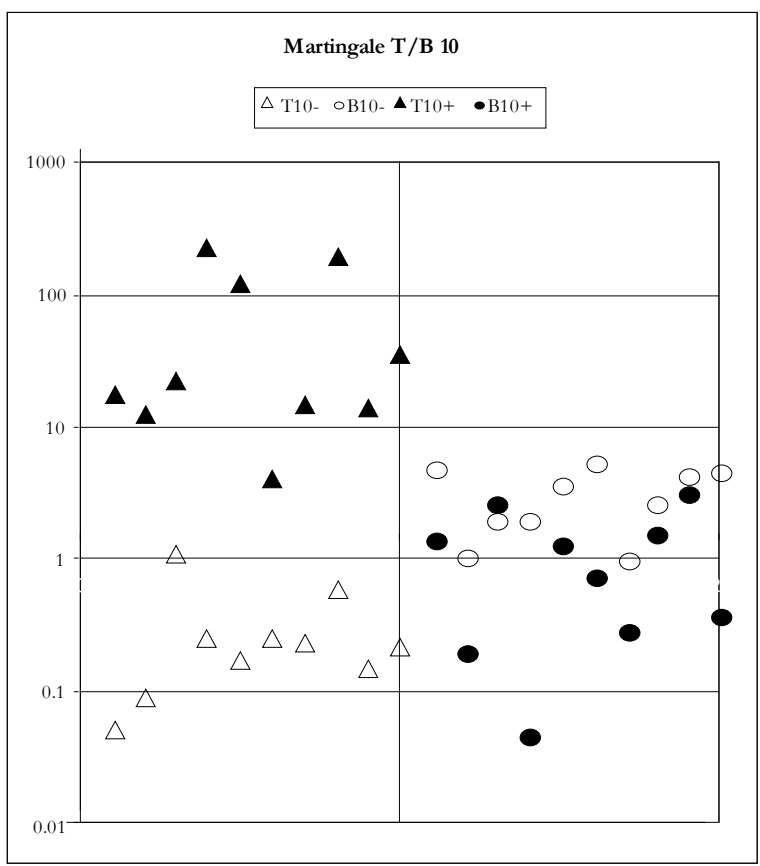

**Figure 12:** Pseudo-compounded returns across all quarters, for Top and Bottom 10 MGL predicted stocks, segregated by the success ("+", full, "in phase") or lack of success ("−", hollow, "out of phase") of the MGL predictor. The x-axis shows the rank (top 1 through 10) of the equity to the left side of the vertical line, and the rank (bottom 1 through 10) on the right side; the y axis the pseudo-compounded returns.

There are four components to **Figure 12,** each with ten differently-coded elements:

▲ T10 MGL portfolio pseudo-compounded return during a (+) "phase" of the predictor
△ T10 MGL portfolio pseudo-compounded return during a (-) "phase" of the predictor
● B10 MGL portfolio pseudo-compounded return during a (+) "phase" of the predictor
○ B10 MGL portfolio pseudo-compounded return during a (-)"phase" of the predictor

The left hand axis indicates the multiple of an initial return of 1.0 at the end of all winning (losing) quarters, were returns to be compounded back to back. ("Pseudo-compounded". Later, we will combine all winning and losing quarters to show that these figures are not unrealistic even though they appear to be when segregated in this way).

Note that full marks represent data aggregated (compounded as described above) from winning quarters; hollow marks represent data aggregated from losing quarters. The left half of the horizontal axis represents equities ranked T1 through T10 across all quarters (from left to right, i.e., ranked 1 through 10), the right half equities B10 through B1 (from left to right, i.e., ranked 1443 through 1452). The leftmost full and hollow marks are in the same horizontal position, likewise for all remaining nineteen successive rightward full and hollow marks. The full leftmost mark represents the pseudo-compounded return for just the number 1 ranked equity during all quarters with the predictor in phase with the market, the hollow leftmost mark for just the number 1 ranked equity during all quarters with the predictor out of phase with the market; the full second-to-leftmost mark represents the pseudo-compounded return for just the number 2 ranked equity during all quarters with the predictor in phase with the market, and so on.

Concentrating first on just the full marks (successful, "in phase" quarters, "$\phi^+$"), note that the marks to left of the midpoint are triangular and represent from left to right the aggregated pseudo-compounded returns respectively for the top 1 through top 10 ranked equities; marks to the right of the midpoint are circular and represent from left to right equities ranked 1443 through 1452 (the bottom ten in order). Each triangle represents the pseudo-compounded final return over 18 continuous quarters. These are quarters during which the MGL predictor generated positive net returns for a T10/B10 hedged portfolio. Although there is very wide scatter on individual real returns, in this aggregated transformed data, in all "in phase" ($\phi^+$) quarters, all top 10 data points lie above all bottom 10 data points.

Turning now to the hollow marks, we find that the above relationship has been largely inverted:

For 11 "out of phase" quarters ("$\phi^-$"), the aggregated data points representing equities predicted to perform as the top 10, are now found entirely below both the full marks representing the top 10 for winning quarters and below the hollow circular marks representing the predicted bottom 10 performing equities. Furthermore, with one exception, each hollow circle lies above its corresponding full circle. That is, during $\phi^-$ quarters, rather than doing poorly, as predicted, the bottom 10 of the MGL model consistently outperform the bottom 10 of the MGL model during winning quarters, often by an order of magnitude or more; as well as greatly outperform the predicted top ten of the MGL predictor itself for those same quarters, and some proportion of the MGL top 10 predictor during winning quarters.

Thus, the MGL predictor—which simply uses the prior actual rank ordering of the market as a predictor—comes very close to demonstrating two distinct states that are evident at a glance—at least at its extremes. (The same phenomenon is evident, if somewhat less clearly, using charts of similarly aggregated and transformed data for the top and bottom 20, 30, …, 100 as well as using non-transformed and non-aggregated data.) With this in view, the $\phi^+$ quarters may be thought of as periods of "positive" predictor state in the sense that the state of the predictor is in phase with the performance of the universe of equities;. Declining quarters we deem periods of "negative" predictor state $\phi^-$.

However, we may observe at least one significant departure from a strict inversion: During periods of negative predictor state $\phi^-$, the positive return of the B10 portfolio is not nearly so great as the positive return of the T10 portfolio during periods of positive state $\phi^+$. In consequence, the MGL predictor generates a net positive return, especially for the T10 portfolio alone (and similarly for T20-T100). We keep this asymmetry in mind as we turn to the ANN portfolio with which we compare it, at the end adding numerical quantification.

**Figure 13** shows the comparable chart for the ANN predictor and the resulting Top and Bottom 10 portfolios:

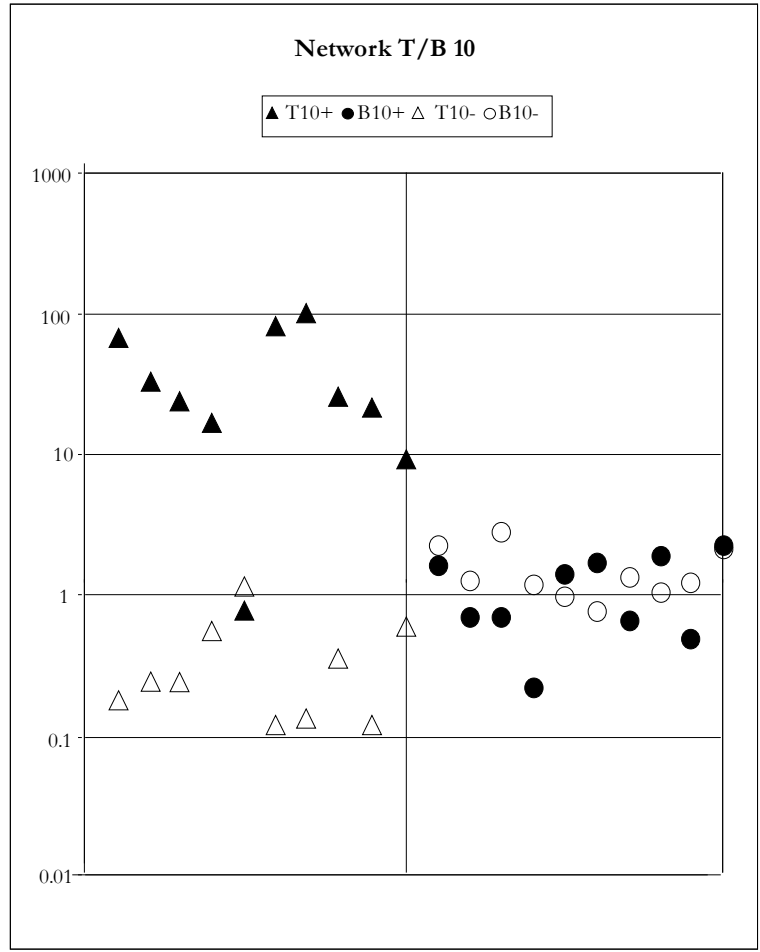

**Figure 13:** Pseudo-compounded returns across all quarters, for Top and Bottom 10 ANN predicted stocks, segregated by the success ("+", full, "in phase") or lack of success ("−", hollow, "out of phase") of the ANN predictor. The x-axis shows the rank (top 1 through 10) of the equity to the left side of the vertical line, and the rank (bottom 1 through 10) on the right side; the y axis the pseudo-compounded returns.

There are likewise four components to the above chart each with ten differently-coded elements:

▲ T10 ANN portfolio pseudo-compounded return during a (+) "phase" of the predictor
∆ T10 ANN portfolio pseudo-compounded return during a (-) "phase" of the predictor
● B10 ANN portfolio pseudo-compounded return during a (+) "phase" of the predictor
○ B10 ANN portfolio pseudo-compounded return during a (-)"phase" of the predictor

These results from the ANN predictor clearly have many general features in common with the MGL predictor. However, there are a number of important differences. Chief among these is the fact that in the right half of the chart (representing the selection of stocks predicted to fall or do poorly, thus suitable for shorting), the circular marks representing the B10 aggregated data from both the positive and negative phases are relatively closely clustered and interpenetrating rather than separated. This is not an artifact of the semi-log representation: In contrast to the MGL predictor, four of ten ANN negative phase B10 lie above their positive phase counterparts. Thus, as a time-averaged statement, the ANN predictor only inverts its prediction for the **top** 10–100 equities during periods of lack of success; it simply loses any predictive ability for the bottom during these phases, which leaves long and hedged strategies intact.

Also, while the ANN predictor does not make as many exceptionally large correct predictions on the positive side, it makes far fewer mistakes in general, especially when selecting equities to sell short (generally a much more difficult task, especially in a rising market).

We may understand the ANN's success vis à vis the MGL predictor as follows. First, we remind ourselves that by "success" we at times simply mean a lesser degree of failure (which for purposes of investment may call forth a high premium). Second, we note that the VL system, the MGL predictor and the top half of the ANN predictor are all subject to an inversion of their behavior; and place this observation in context of the above observations: The possibility that when they fail, the simpler models are falling prey to a change occurring in the market as a whole which they cannot anticipate or adjust for.

Third, in both the MGL and ANN predictors, inversions most often occur *delayed* relative to changes in the universe of equities, but this happens less often with the ANN. (Hence the anti-persistence evident in both as discussed above, but the greater anti-persistence for the MGL.) This makes sense, of course, as it explains why a large number of losing quarters arise—e.g., 9 of 29 for the H10 ANN predictor, 11 of 29 for the H10 MGL predictor, in the statistics provided above.

Studying this problem carefully requires a more rigorous definition of what constitutes a "state" than we have constructed, and what constitutes an inversion. However, a quick impression can be obtained from **Figure 14**:

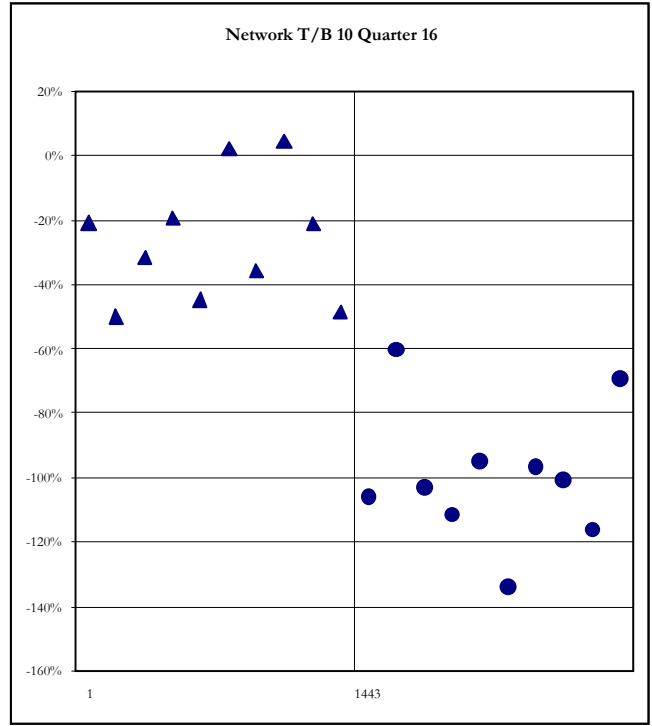

**Figure 14:** ▲ Top 10 ● Bottom 10
Quarter 16 T10 and B10 equities ranked by the ANN within full rank spectrum of 1443 stocks (of 1452 possible).

**Figure 14** shows the percent returns for the top and bottom 10 equities ranked by the ANN just for quarter 16 showing their actual ranks within all stocks available during that quarter, during which the universe of (during that quarter) 1443 (out of the possible 1452) equities experienced a mean decline of 21.8%. (There are usually fewer stocks to rank within a given quarter because of the listing and delisting of corporations.) The preceding quarter saw a rise of 2.7% and the following quarter a rise of 24.3%. During this quarter, for this portfolio, the MGL predictor persisted in maintaining the "usual" ranking structure which had a (weakly) inverse relationship to the actual market during this quarter. While all four of the T10, B10 MGL and ANN *individual* portfolios lost—T10 MGL, B10 MGL, T10 ANN and B10 ANN—the T10 MGL did *worse* than the B10 MGL for a net H10 MGL loss during this one quarter of almost 4%.

On the other hand, while the ANN predictor does not resist the overall downward trend of the market and even—consistent with its overall volatility—amplifies it, the ANN H10 portfolio makes a significant gain. The ANN has created a winning rank-ordering structure for a hedged portfolio almost entirely from negative returns, in this case, in the face of (or composed out of) a declining market (following a series of quarters in which the overall market had risen, but less so, in each successive quarter). Note that in the long run, successful results are obtained for the T10-T100 portfolios on an absolute basis.

In the additional two quarters where the ANN succeeds and the MGL predictor fails, the market as a whole was similarly experiencing an overall decline (as in the one example just given). The successful adaptation in the structure of the predictor occurs "simultaneously" with the change in the equity market as a whole (within the discrete time unit of the procedure—one quarter).

The ANN predictor resists simply following the prior state, and rather seems to adapt when called for (with mixed success), with less of a delay. Hence, there are quarters when instead of failing because it does not keep up with a changing market, it partially adapts. The ANN predictor is much more frequently able than the MGL predictor to generate a ranking structure that is useful during generally adverse periods.

**Table 5** and

**Table** 6 present descriptive statistics comparing the MGL and ANN predictors for the T10 and B10 combined portfolios using the above aggregated measures. The figures in **Table 5** present *idealized* mean quarterly returns that would yield the pseudo-compounded returns in

**Table** 6. Note that whereas the idealized pseudo-compounded ANN returns are approximately equal to the actual compounded returns in the simulation, the idealized MGL returns are much larger. Thus fortunately, this comparison has tended conservatively (in this instance) to favor the MGL predictor.

The different results arise as follows: The product of 1 plus each specific quarterly difference between T10 and B10 is not equal to the product of 1 plus their average difference. I.e., in general, "pseudo-compounded" returns calculated as $\left(1+\frac{1}{n}\sum_{i=1}^{n}x_i\right)^n = (1+\bar{x})^n \neq \prod_{i=}^{n}(1+x_i)$, the last expression being the actual cumulative returns. For example, if one of the specific factors $x_j = -1$, the final product $\prod_{i=}^{n}(1+x_i) = 0$, whereas the second product would equal zero only if $\frac{1}{n}\sum_{i=1}^{n}x_i = \bar{x} = -1$.

The data in

**Table** 6 present the data represented in **Figure 13**. Together they quantify the preceding discussion.

**Table 5:** Mean quarterly returns segregated by "$\phi$"

| Quarterly Mean | $\phi^+$ | $\phi^-$ |
|---|---|---|
| *No. Quarters* | 20 | 9 |
| ANN T10 | 0.20 | -0.106 |
| ANN B10 | 0.01 | 0.052 |
| **ANN H10** | **0.19** | **-0.16** |
| *No. Quarters* | 18 | 11 |
| MGL T10 | 0.26 | -0.10 |
| MGL B10 | 0.01 | 0.11 |
| **MGL H10** | **0.25** | **-0.21** |

**Table 6:** Pseudo-compounded quarterly returns segregated by "$\phi$"

| Ps.Compounded | $\phi^+$ | $\phi^-$ | Cum. ret. | Ann. ret. |
|---|---|---|---|---|
| *No. Quarters* | 20 | 9 | | |
| ANN T10 | 38.55 | 0.37 | | |
| ANN B10 | 1.22 | 1.58 | | |
| **ANN H10** | **32.72** | **0.21** | **6.98** | **30.7%** |
| *No. Quarters* | 18 | 11 | | |
| MGL T10 | 67.21 | 0.31 | | |
| MGL B10 | 1.19 | 3.16 | | |
| **MGL H10** | **58.64** | **0.07** | **4.29** | **22.3%** |

During 20 "positive" regime quarters ("$\phi^+$": predictor in phase with the universe of equities), the ANN predictor generated average returns of 0.191 for the H10 portfolio (combined T10 and B10). For 9 "negative" regime quarters ("$\phi^-$": predictor out of phase with the universe of equities), the ANN predictor generated average returns of −0.158 for the H10. By contrast, the MGL predictor generated larger mean positive regime returns of 0.254 for the H10, but for fewer quarters (18) and also generated larger negative regime returns of −0.212 for more quarters (11).

**Table** 6 illustrates the equivalent pseudo-compounded results with 30.7% annualized return for the ANN predictor (close to actual) and 22.3% for the MGL (significantly higher than actual).

The most important point to be noted is that the relative success of the network predictor—and relative failure of the MGL predictor—is caused by the relatively large values in the negative phase columns, MGL B10 rows, compared to the ANN B10 rows, as these values are contrasted to their respective T10 rows.

### 5.4.2 T/B 100 = H100 Portfolios

The above discussion can be extended throughout all the portfolios from 10 through 100 with (almost) uniform results. **Figure 15** and **Figure 16** are comparable to **Figure 12** and **Figure 13**; **Table 7** and **Table 8** are comparable to **Table 5** and

**Table** 6, presenting results for the T100, B100 and H100 portfolios, both ANN and MGL.

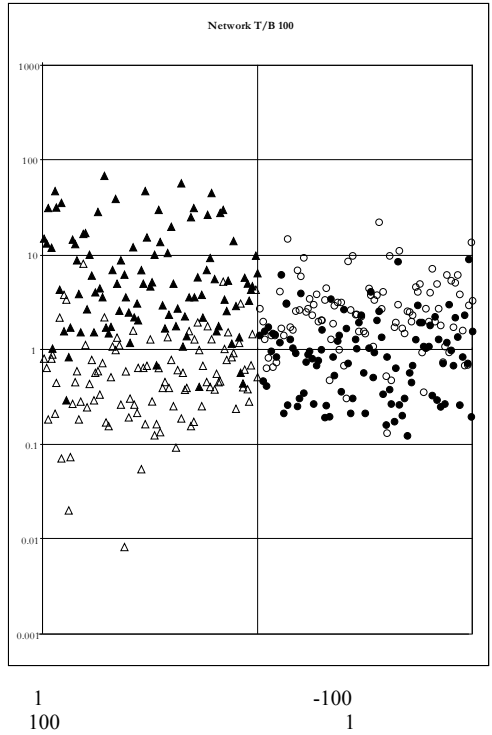
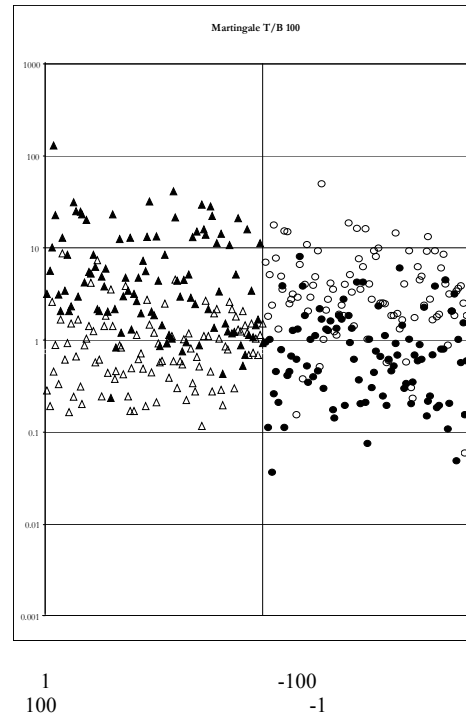

**Figure 15**: ANN T/B = H100 Portfolios     **Figure 16**: MGL T/B = H100 portfolios

**Table 7:** Mean quarterly returns segregated by phase $\phi$

| Quarterly Mean | $\phi^+$ | $\phi^-$ |
|---|---|---|
| *No. Quarters* | *16* | *13* |
| ANN T100 | 0.20 | 0.01 |
| ANN B100 | 0.03 | 0.10 |
| **ANN H10** | **0.17** | **−0.09** |
| *No. Quarters* | *15* | *14* |
| MGL T100 | 0.19 | 0.036 |
| MGL B100 | 0.01 | 0.163 |
| **MGL H100** | **0.18** | **−0.13** |

**Table 8:** Pseudo-compounded quarterly returns segregated by phase $\phi$

| Ps.Compounded | $\phi^+$ | $\phi^-$ | Cum. ret. | Ann. ret. |
|---|---|---|---|---|
| *No. Quarters* | *16* | *13* | | |
| ANN T100 | 14.91 | 1.11 | | |
| ANN B100 | 1.58 | 3.90 | | |
| **ANN H100** | **12.50** | **0.28** | **3.49** | **18.8%** |
| *No. Quarters* | *15* | *14* | | |
| MGL T100 | 11.88 | 1.68 | | |
| MGL B100 | 1.10 | 8.80 | | |
| **MGL H100** | **11.56** | **0.15** | **1.72** | **7.8%** |

Here again the same general pattern prevails, if somewhat attenuated. The superior relative and absolute performance of the ANN predictor is attributable primarily to its capacity to occasionally adapt rank structure in phase with shifts in the market. It therefore has a larger number of winning quarters. Consistent with this capacity, it tends to "resist" an apparent inversion at the bottom end of the ranking during losing quarters, thus incurring smaller losses than the MGL predictor.

# 6 Discussion

The prediction method outlined above has many complex features only a few of which have been treated. We discuss here a small number of key points, including problems.

## *6.1 Data Cleaning and Errors*

As discussed previously, the original VL data has an extremely large number of errors. It was not possible to obtain meaningful results using uncleaned data and the process of cleaning data to an acceptable level is very resource-intensive. Because of this, it has not yet been possible to run simulations on all possible monthly, weekly and daily cycles within a quarter. Randomly selected and targeted runs performed on partial sets show insignificant differences with the results presented here. Identical simulations performed at shorter data intervals (monthly, weekly, daily) should yield similar results with more robust statistics. But as the interval shortens, the data problem worsens. Furthermore, with respect to earnings, the quarterly time period is the only natural time step.

## *6.2 Why Does the ANN Predictor Work?*

There are a sufficient number of occasions when the ANN predictor seems able to adapt and anticipate an inversion in the market, i.e., a change in the structure of the VL universe of equities it is attempting to rank, and so undergoes a change in the rank-ordering it establishes. Rather than mimic the rank-ordering of the immediately prior quarter (as does the MGL predictor), the ANN has the option of looking back at the rankings by change in price and by change in earnings for all 1452 equities over ten quarters. For each equity it then assesses the impact of each of these twenty parameters on the most recent change in price. But every other input is itself an earlier period's change in price (merely transformed into a rank), that was itself once an output. Not every quarter most closely mimics the prior quarter. There are in fact many quarters when the average change in price over the entire universe of equities more closely mimics what happened two quarters prior, or three. (A MGL predictor based on the rank ordering of two quarters prior still outperforms a random ranking.) To some extent, the ANN is apparently able to properly weight the changing relative contribution of these prior quarters, including the contribution of changes in earnings: An identical predictor that excludes earnings inputs still outperforms the MGL predictor, but underperforms this ANN predictor and is much more volatile. In particular, because the ANN tacitly develops a nonlinear relationship among the input variables, it may detect as a "pattern", a pattern in the "change-of-state" of the market, as it were, if such a pattern exists or seems to exist.

This, of course, raises a caution: With only 29 quarters, it is not possible to differentiate between a "pattern of change of state" that actually exists and one that only seems to exist, even if the internal structures within each quarter are very robust. In other words, the ANN predictor may indeed be better at detecting global patterns of the above kind than the MGL predictor. But one can only *argue* that the markets undergoing large scale declines during the quarters captured by the ANN (and missed by the MGL) did so as part of a pattern detectable by the ANN rather than a pattern invented by it. The success of the ANN in generating large returns provides the evidence for the former.

### *6.3 Relation between the ANN and the VL ranking system*

The behavior of the ANN in generating rankings shows many similarities to the VL system proper which, as noted in the introduction, has been the object of significant attention because of its apparently anomalous success. One similarity is the ability of the ANN to generate significant positive returns; another is the fact that when it fails, it, too, does not simply lose predictive ability, it produces rank-orderings that are inverted relative to the actual price changes and so creates large losses. Nonetheless, better than a MGL predictor, or the VL system proper, it is able both to anticipate the need for a different rank ordering, on occasion, and to do a better job in resisting failure when identifying stocks for short-selling. On balance, at least in this study sample, the ANN is able to improve upon the VL approach proper and generate net positive returns over the long run in excess of a buy and hold strategy and sufficient to overcome transaction costs. (These are kept to a minimum in the quarterly trading process employed). These results suggest that at least part of the power inherent in the VL approach is in wide use of rank orderings as a general method for coarse-graining financial data.